# Advancing NASA-TLX: Automatic User Interaction Analysis for Workload Evaluation in XR Scenarios

1st Aida Vidal-Balea
Dept. of Computer Engineering
CEMI-Centro Mixto de Investigación UDC-Navantia
Ferrol, Spain
aida.vidal@udc.es

2nd Paula Fraga-Lamas
Dept. of Computer Engineering
CITIC Research Center, UDC
A Coruña, Spain
paula.fraga@udc.es

3rd Tiago M. Fernández-Caramés
Dept. of Computer Engineering
CITIC Research Center, UDC
A Coruña, Spain
tiago.fernandez@udc.es

*Abstract*—Calculating the effort required to complete a task has always been somewhat difficult, as it depends on each person and becomes very subjective. For this reason, different methodologies were developed to try to standardize these procedures. This article addresses some of the problems that arise when applying NASA-Task Load Index (NASA-TLX), a methodology to calculate the mental workload of tasks performed in industrial environments. In addition, an improvement of this methodology is proposed to adapt it to the new times and to emerging Extended Reality (XR) technologies. Finally, a system is proposed for automatic collection of user performance metrics, providing an autonomous method that collects this information and does not depend on the users' willingness to fill in a feedback questionnaire.

## I. INTRODUCTION

The arrival of cutting-edge technologies designed to improve production processes has driven unprecedented growth, setting the stage for what is widely recognized as the fourth industrial revolution. Consequently, the term Industry 4.0 was recently introduced to describe this transformative era, in which industrial operations seamlessly integrate manufacturing processes, information systems and communications technologies. Specifically, Industrial Internet of Things (IIoT) and Industrial Augmented Reality (IAR) stand out as key components within the Industry 4.0 paradigm [1].

When using Extended Reality (XR) technologies, which include IAR and Industrial Mixed Reality (IMR), effective application evaluation is essential to ensure optimized user experiences. This paper presents research that merges the recognized NASA-Task Load Index (NASA-TLX) methodology [2] with an innovative way to automatically collect usage data in order to evaluate mental workload and user experience within XR applications.

This work has been funded by grant PID2020-118857RA-100 (ORBALLO) funded by MCIN/AEI/10.13039/501100011033 and by Centro Mixto de Investigación UDC-NAVANTIA (IN853C 2022/01), funded by GAIN (Xunta de Galicia) and ERDF Galicia 2021-2027.

In this paper, a first approach for XR NASA-TLX is proposed, adapting the NASA-TLX methodology to address the specific challenges of XR. This new proposed system automatically collects user performance metrics, and has the aim of measuring how easy or difficult it is to perform a task in industrial workshops using smart glasses as a substitute for the traditional techniques used by operators.

In order to create an automatic system for collecting usage data, an application for Microsoft HoloLens 2 smart glasses was developed. The selected use case consisted in an application for an Industry 4.0 environment aimed at speeding up and improving the efficiency of the operators in the tasks of laying and installing the electrical boilerwork during the construction process of a ship.

## II. STATE OF THE ART

A range of tools for mental workload assessment and prediction can be found in the literature, like NASA-TLX [2], Subjective Assessment Technique (SWAT) [3] or Workload Profile (WP) [4], which stand out as methodologies that measure the workload of a task in a subjective way.

The first methodology, NASA-TLX, uses six dimensions to assess mental workload. Before performing the task, each dimension is first weighted, making a binary comparison between all the dimensions and selecting the most relevant dimension to workload across all pairs of the six items. Then, after the task is completed, each dimension is assessed using twenty-step scales, with ratings obtained on a scale of 0 to 100. A variation of NASA-TLX is SIM-TLX [5], which was developed specifically for assessing Virtual Reality (VR) simulations. Such a new methodology makes use of a total of 10 dimensions and is also split into two phases.

The next methodology, SWAT, makes use of three dimensions: time, mental effort and psychological stress; as well as three levels: low, medium, and high (rated from 1 to 3). All the process is divided into three different steps: firstly, the scale development phase involves creating 27 cards representing all possible combinations of three levels across three dimensions. Operators then rank these cards according to their perceived increase in workload. The second step is the event-scoring phase, where actual workload ratings are assigned to specific tasks. Finally, in the third step, the ratings for each dimension are converted into numeric scores ranging from 0 to 100, utilizing the interval scale established in the initial phase.

The last methodology, WP, uses four dimensions, each divided in two, giving a total number of eight subdimensions. After all the tasks are completed, users are asked to give a value between 0 and 1 for each subdimension (being 0 a low-demand and 1 a high-demand subdimension), repeating this procedure for each completed task.

A study was found where these three methodologies, NASA-TLX, SWAT and WP, were compared [6], pointing out that the dimensions used by WP generated some comprehension problems and concluding that if the goal is to predict the performance of a particular individual in a task, then NASA-TLX is recommended.

Therefore, analyzing the previously mentioned alternatives in combination with the comparison between the complexity of the questionnaires of the different methodologies, it was concluded that NASA-TLX was the most appropriate when calculating the mental workload of a task.

Regarding SIM-TLX, although it is useful for VR, it is strongly oriented towards the evaluation of simulated tasks, while in Augmented Reality (AR) and Mixed Reality (MR) the environment is not simulated because the experience is built on top of the real world. It should also be noted that SIM-TLX uses a total of ten dimensions: mental, physical and temporal demands; frustration; task complexity, situational stress, distraction, perceptual strain, task control and presence [5]. The problem encountered is that having 10 different dimensions that are evaluated at the same level, leads to an increase in the number of binary comparisons to be performed. This causes going from the 15 comparisons carried out by NASA-TLX to the 45 required by SIM-TLX, making the weighing phase of the questionnaire excessively long. In addition, as it was previously mentioned, using dimensions as presence, task control and distraction causes the SIM-TLX methodology to not be directly applicable to all XR scenarios, since such dimensions are aimed at evaluating task load in immersive VR environments.

Due to the previous reasons, it was decided to design a methodology based on the original NASA-TLX that can be applied to any XR experience, not only to simulated ones, minimizing at the same time the complexity of the forms that the users have to fill in.

## III. FROM NASA-TLX TO XR NASA-TLX

NASA-TLX is one of the best known tools for assessing the subjective workload required by a task. This methodology provides a multidimensional assessment procedure that gives an overall workload score based on a weighted average of scores on six subscales or dimensions: mental, physical and temporal demands; performance; effort; and frustration level [2].

The methodology is divided into two phases: weighting and scoring [7]. The first phase, carried out prior to the execution of the task, allows users to weight the sources of mental workload, in order to determine the degree to which each of the six factors affects the evaluated task. Therefore, the weighting phase consists of a pairwise comparison (binary comparisons) in which users choose, for each pair, which element has a greater mental workload. An example of this weighting can be seen in Fig. 1. After completing the weighting phase, a value is obtained for each dimension, where the most frequently selected items are given the highest weighting. The second phase is conducted immediately after the task has been completed. This phase consists of asking users to rate the task that has just been performed, on a scale split into twenty segments, for each of the six dimensions mentioned above. An example of the score requested to the users can be seen in Fig. 2.

Fig. 1. Screenshot of the proposed XR NASA-TLX: Phase 1.

Fig. 2. Screenshot of the proposed XR NASA-TLX: Phase 2.

As it was previously mentioned, the NASA-TLX methodology has been widely used to assess mental workload in a variety of tasks [8], but its applicability is not fully developed when it comes to dealing with emerging technologies such as XR. Although it is still a valid methodology for assessing mental workload, NASA-TLX was created in the late 1980s, when XR technologies were not as popular as today. Therefore, in order to fill this gap in the NASA-TLX methodology, this paper proposes XR NASA-TLX, an adaptation and expansion of the traditional methodology that addresses more precisely the assessment of mental workload in tasks performed using XR. To this end, five new dimensions that take into account specific aspects related to the use of these technologies are designed and added, with the aim of improving the sensitivity of the workload evaluation:

- *ĭ* **Physical comfort** (high/low): Are the glasses comfortable to wear or do you experience any physical discomfort? (e.g. headache, excessive weight, etc.)
- *ĭ* **Visual comfort** (high/low): Is it comfortable to see the objects projected by the glasses or do you experience any discomfort? (e.g., eye discomfort or irritation, field of view, image sharpness, etc.)
- *ĭ* **General comfort** (high/low): Overall, are the glasses comfortable to wear or do you experience any discomfort? (e.g., dizziness, disorientation, loss of balance, etc.)
- *ĭ* **Ease of use** (high/low): Is the application easy to use? Is it intuitive? Are the menus well understood? Are the necessary items found quickly?
- *ĭ* **Application usability** (high/low): Do you consider that the application is useful as a substitute for paper blueprints?

The aim behind adding these five new items in the surveys is to collect information related to the used technology and thus be able to assess whether it helps or interferes in any way in the performance of the task or implies a greater mental workload. The dimensions include both the comfort aspect of wearing the glasses and the technological aspect by asking questions about the application itself. The inclusion of these aspects in relation to the technology was inspired by the research carried out by Noyes and Bruneau [8], who compared the workload of carrying out the NASA-TLX methodology by covering the forms on paper and on a computer. They concluded that, for their case scenario, the use of computers was counterproductive, since they required a greater effort from the users.

The final XR NASA-TLX questionnaires are shown in Figs. 1 to 3. In such forms, the panels labeled as "task" are those referring to the traditional NASA-TLX dimensions, while those labeled as "technology" are part of the new XR NASA-TLX methodology. The initial part of the form, shown in Fig. 3, is the introductory section, where the different dimensions to be evaluated are detailed and explained. After such a form are phases 1 and 2 of the XR NASA-TLX methodology, whose questionnaires are shown in Figs. 1 and 2, respectively.

## IV. Experiments

### A. Practical use case

In order to test whether the proposed XR NASA-TLX methodology was accurate, an industrial practical use case from Centro Mixto de Investigación UDC-Navantia (CEMI) was chosen. This project was carried out in collaboration with Navantia's shipyard in Ferrol, in which an industrial application of MR was developed with the main objective of speeding up and improving the efficiency of operators in the tasks of placement and installation of electrical boilermaking elements during the shipbuilding process. Fig. 4 shows a screenshot of the application where some main elements can be seen: the vessel's structure along with the electrical boilermaking elements (supports and through-bolts) that have to be welded by the operators in the workshops.

This application can place overlapping 3D boilermaking elements and their corresponding dimensions directly on the vessel being built, as shown in Fig. 5, so that they are perfectly aligned and can be used as a reference by the technicians to take measurements and know in which exact position the boilermaking elements have to be welded, without the need to read and interpret the traditional 2D blueprints.

All experiments, which are detailed in Subsections IV-B to IV-D, were performed making use of the previously described XR application. The aim of these experiments was to perform a preliminary evaluation of the XR NASA-TLX methodology and check whether this adaptation is on the right track to become a useful tool for measuring the mental workload of tasks executed under XR technologies.

Fig. 3. Screenshot of the proposed XR NASA-TLX: introduction.

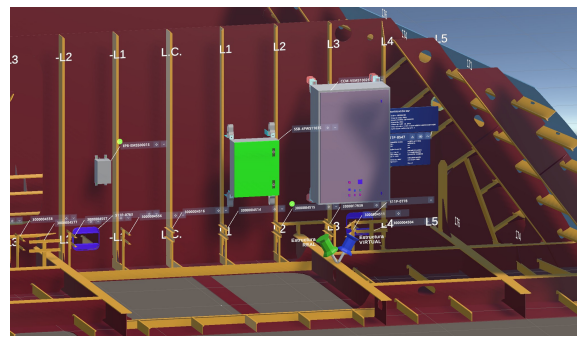

Fig. 4. General 3D view of the application: vessel structure and boilerwork elements (supports and through-bolts) to be placed by the operators.

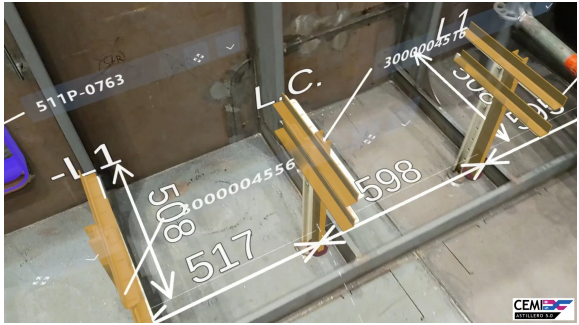

Fig. 5. Detail of multiple supports (virtual element overlapped with the real one) and corresponding dimensions, as seen from Microsoft HoloLens glasses.

*B. First experiment: Testing XR-NASA-TLX adaptation*

During the first phase of the experiments, the proposed XR NASA-TLX methodology was applied in order to compare the mental workload required to perform a task in the traditional way versus the same task when using the developed MR application.

For this purpose, 16 shipyard workers, both men and women, collaborated in this process. They were randomly divided into two groups of 8 people. Both groups were asked to perform two tasks: the first one consisted of finding a boilermaking element inside the ship, and the second task was to locate the point at which an element had to be welded. In both cases, the element identifier number was provided. The first group was asked to perform the tasks in the traditional way, using 2D blueprints, whereas the second group would perform the tasks using the MR application. Thus, of these people, about 50% had plenty of experience performing the assigned tasks and only a few had used the MR smart glasses beforehand.

Prior to performing the tasks, the different phases of the XR NASA-TLX methodology were described to the attendees, as well as how the forms for phases 1 and 2 had to be filled in. It was also explained what the tasks they had to perform consisted of and it was shown in real time how to use the MR application. Once in the workshop, after one worker performs the assigned tasks, the paper forms distributed at the beginning were collected.

At the end of the experiment, the total number of collected forms was 9 (3 forms from the first group and 6 from the second group), of which 2 were blank and 1 was incomplete. Finally, of the 16 forms initially distributed, data could only be retrieved from 6 forms, 2 from the first group and 4 from the second.

The first surprising fact when comparing the results obtained for the mental workload of each of the tasks is the disparity of the results. However, this is to be expected, since, as mentioned above, XR NASA-TLX is a subjective methodology, and therefore it is influenced by the perceived experience of each user. For example, the two users from the first group have scores of 55-54 and 13-22 for their tasks (maximum difference of 9 points between tasks for the same person), and both of them had the same experience performing the assigned tasks. In contrast, the second group have scores ranging from 33-42 points for the first task and 27-44 for the second (with a maximum difference of 7 points).

In view of the above and and seeing the difficulties that arise when filling in the forms, it was decided that it would be a better idea to incorporate a system for the automatic collection of usage metrics within the MR application itself. This would avoid relying on the users' willingness to complete the data for the XR NASA-TLX surveys.

*C. Second experiment: Automatic user interaction data collection*

As it was previously explained, based on the problems in obtaining results from the XR NASA-TLX forms, it was considered to add a system for the automatic collection of user interaction data. For this purpose, click events and eye-gaze collisions with objects are captured making use of triggers that are fired when interacting with the 3D environment through clicks and MRTK3, which provides APIs to manage collisions between the user's gaze and the objects that integrate the scene. For each of these events, the information that is stored in the database is: date, start time (as well as end time in the case of gaze) and the object that is being interacted with.

By systematically saving these data, the following information can be retrieved: total number of interactions (broken down into clicks and gazes), application usage time, average number of gazes per minute as well as clicks per minute and total number of objects that have been paid attention to (those objects that have been looked at for one second or more).

Once the automatic system for usage data collection was developed, the second experiment could take place. Again, it was carried out in collaboration with shipyard operators. To this end, a group of 23 people, both workshop workers and engineers, was gathered. The only information that was asked to the testers was their level of experience with the technology (in this case with the MR HoloLens glasses) and whether they knew or were familiar with the application beforehand. To perform the experiment, the testers were requested to try out the application and interact freely with it and their environment. In order to try to intervene as little as possible, a brief introduction was given only to those people who did not yet know the application or did not know how to use the glasses. For this purpose, a video showing the menus and the basic features of the application was projected.

Below are the conclusions drawn from the detailed analysis of the data obtained during the experiment. The following graphs illustrate how the level of knowledge of the application and the experience level with the Microsoft Hololens glasses relate to the usage metrics.

One of the first things that were analyzed was the number of interactions of each user. Fig. 6 shows a trend indicating that the higher the level of knowledge of the application, the higher the number of interactions. This suggests that users with deeper knowledge tend to interact more with the application. Although it should be noted that it is conditioned by the time of use, since it is highly probable that the longer the time of use, the higher the number of total interactions are performed.

Likewise, it was also considered important to examine how many objects each user interacts with visually (i.e., how many objects the user focuses or pays attention to). For the evaluated case, it was decided that it would be considered an interaction of interest if the user looks at an object for at least 1 second. Fig. 7 shows that the level of knowledge of the application positively influences the number of objects that are paid attention to. This indicates that users who are more familiar with the application tend to pay more attention to a higher number of objects, possibly because they explore and interact with the environment more deeply.

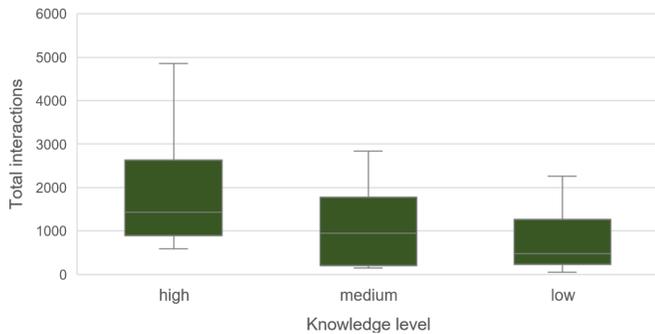

Fig. 6. Number of total interactions, users sorted by level of knowledge of the application (high, medium, low).

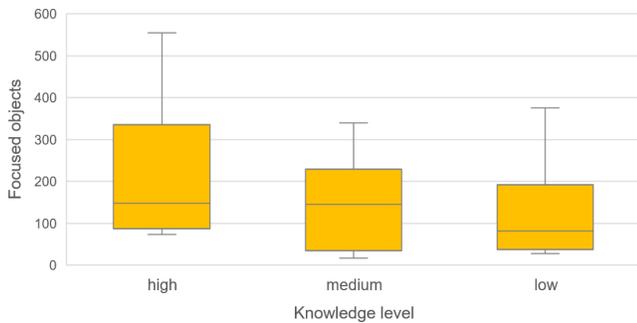

Fig. 7. Number of focused objects, users sorted by level of knowledge of the application (high, medium, low).

The comparison between users who were and were not previously familiar with the technology, here considered as "expert users" and "non-expert users" respectively, is shown in Fig. 8, where expert users tend to perform more clicks per minute. This may indicate greater efficiency or familiarity with the interface, as well as an extra comfort with the environment in which the users operate.

In addition, the number of focused objects (objects observed for 1 second or longer) also appears to be higher in users who already knew how to use the Hololens, as it can be seen in Fig. 9. As indicated above, this suggests greater attention or interaction with the application by these users.

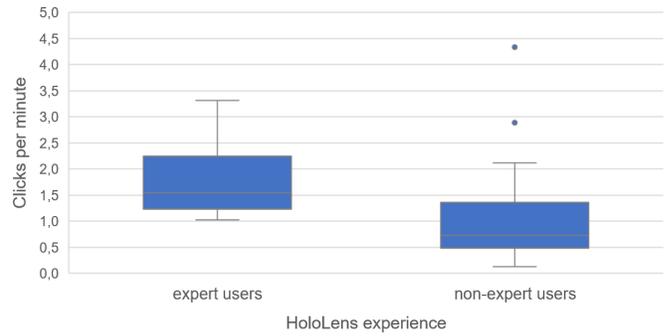

Fig. 8. Number of clicks per minute, users divided by level of previous HoloLens experience (high/low-none).

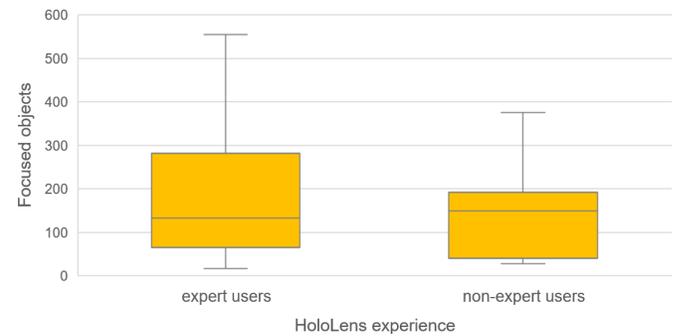

Fig. 9. Number of focused objects, users divided by level of previous HoloLens experience (high/low-none).

### D. Third experiment: comparing technology and application knowledge levels

Finally, a third experiment was carried out in order to compare three users with different levels of experience with MR technology and also different degrees of knowledge of the application.

The first user had both a low level of experience with HoloLens and a low familiarity level with the application. The second user had a high level of experience with the smart glasses and a medium level of familiarity with the application. Finally, user number three had both a high level with the smart glasses and with the application.

All three users were asked to use the MR application for completing two tasks, both very similar to those performed during the first experiments carried out, explained in Sec. IV-B. The first task consisted in searching an electrical boilermaking element, for which the identifier was provided, and then generating a notification indicating that there was a problem and that the item was damaged. The second task involved welding an element. Once again the identification code was provided, and the user had to find the element and mark the real exact point where it would be welded. In contrast to the first experiment, these users had none to hardly any experience performing the given tasks in the shipyard, but they differed in their knowledge level of the application.

The processed data gathered by the system are shown in Fig. 10. One of the things that can be easily noticed is that the gaze per minute is quite regular among the three individuals, decreasing slightly as the users' knowledge of the application increases. This fact is possibly due to them knowing where each of the functionalities is located, so there is no need to search within the application. On the other hand, as it was previously mentioned, total interactions and focused objects are directly proportional to the time of use. Moreover, the difference in time and efficiency (clicks per minute) in the use of the application is remarkable comparing the user with less and more experience. As for user number two, it is worth noting that, after the completion of the tests, he reported that during the first task he got confused when searching for the item, and therefore his total usage time and other metrics are higher than in the two other cases.

Therefore, it can be said that data automatically collected during testing provides evidence that both the level of knowledge of the application and prior experience positively influence the manner and efficiency with which the application is used.

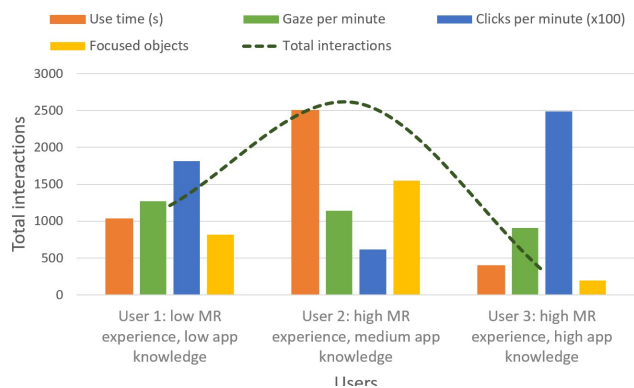

Fig. 10. Metrics comparison between three users performing two given tasks.

## V. CONCLUSIONS

The proposed XR NASA-TLX, an adaptation of the NASA-TLX methodology, includes the incorporation of five new dimensions specific to XR technologies: physical, visual and general comfort, ease of use and application usability. The addition of these new elements to the methodology is intended to improve the ability to assess mental workload in environments involving these emerging technologies by providing a more complete and up-to-date perspective. These modifications seek to address the current limitations of the methodology and ensure its relevance in constantly evolving technological contexts.

Three experiments were performed using an MR application developed during one of the projects carried out in collaboration with Navantia's shipyard in Ferrol. Such MR application facilitates and simplifies the welding processes that workshop operators carry out on a daily basis. For the experiments, a group of individuals were asked to perform two tasks in the workshops. During the first experiment, it was concluded that data collection heavily relied on the willingness of the users, which, due to the inherent structure of the forms, became a rather tedious task. Therefore, it was decided to create a system that automated the extraction of usage data for later evaluation, allowing to track user's performance while completing the tasks using the MR application.

The developed system allows for the automatic collection of performance metrics, enabling it to obtain information on user performance indicators: total number of interactions, average amount of clicks per minute, total number of objects that have been paid attention to (those objects that have been looked at for one second or more), and time of use of the application. In addition, this also provides an autonomous system that does not depend on the user's willingness to provide feedback on the experience of performing a task aided by XR tools.

As future work, on the one hand, it is planned to carry out an exhaustive evaluation of the proposed methodology (i.e., to perform a psychometric evaluation, in order to correctly validate XR NASA-TLX as a tool to accurately measure mental workload for XR tasks). On the other hand, experiments will continue to be carried out in order to find the appropriate metrics that will provide all the necessary data to be able to correctly and automatically evaluate the experience of XR tool users.

## ACKNOWLEDGMENT

We would like to thank all the Navantia personnel who collaborated with us both during the development of the electrical boilermaking MR project and also during the tests carried out in the shipyard workshops. We would also like to extend our recognition to Oscar Blanco and Antón Valladares, researchers at CEMI who collaborated during the development process of the MR application.